# Surface reconstructions and transport of epitaxial PtLuSb (001) thin films grown by MBE

Sahil J. Patel[a], John A. Logan[a], Sean D. Harrington[a], Brian D. Schultz[b], Chris J. Palmstrøm[a,b*]

[a]Materials Department, University of California-Santa Barbara,
Santa Barbara, 93106-5050, United States of America
[b]Department of Electrical and Computer Engineering, University of California-Santa Barbara,
Santa Barbara, 93106-9560, United States of America

(Dated November 13, 2015)

**Abstract**

This work presents the surface reconstructions and transport properties of the topological insulator PtLuSb grown on $Al_{0.1}In_{0.9}Sb$/GaAs (001). Two stable surface reconstructions, (1x3) and c(2x2), were observed on PtLuSb (001) surfaces. Antimony-dimerization was determined to be the nature of the (1x3) surface reconstruction as evidenced by chemical binding energy shifts in the antimony 4d core-level for surface bonding components. The two surface reconstructions were studied as a function of $Sb_4$ overpressure and substrate temperature to create a reconstruction phase diagram. From this reconstruction phase diagram, a growth window from 320 °C to 380 °C using an antimony overpressure was identified. Within this window, the highest quality films were grown at a growth temperature of 380 °C. These films exhibited lower p-type carrier concentrations as well as relatively high hole mobilities.

## 1. Introduction

Heusler compounds are ternary intermetallics, which, depending on the chemical species and ordering, can exhibit novel electronic structures ranging from half-metallic ferromagnets[1–3] to semiconductors[3,4] to superconductors[5]. Despite the wide range of properties, Heusler compounds only form in a few related crystal structures, including the $L2_1$, B2, $DO_3$, and $C1_b$. In addition, these materials are closely lattice matched to III-V semiconductors, enabling integration into existing device technologies. Recently, the existence of topological insulators in the family of half-Heusler compounds has been proposed[6–8]. These materials have an insulating bulk band gap combined with conductive surface states where the electron spin is locked to the momentum. Recent experiments have reported the first direct experimental observation of these topologically non-trivial surface states in Heusler compounds in the topological insulator candidate, PtLuSb[9]. Although there has been significant progress with bulk prepared samples[10], including record mobilities for PtLuSb[11], to utilize and modulate these spin-locked surface states, gateable high quality thin film device structures must be fabricated. During thin film growth, crystalline quality, including bulk defects and surface morphology, is determined by the kinetics of the growing surface; therefore, understanding surface bonding is crucial in determining ideal growth conditions. This work presents the stable surface reconstructions of PtLuSb (001) surfaces and demonstrates how they relate to finding conditions for stoichiometric growth.

## 2. Experimental Methods

PtLuSb (001) films were grown by molecular beam epitaxy (MBE) on relaxed $Al_{0.1}In_{0.9}Sb$ films grown on GaAs (001) substrates. MBE has produced high quality half-Heusler films[12–15] as well as offers the ability to control elemental fluxes and abruptly terminate surfaces. Additionally, MBE provides sub-monolayer deposition control over elemental species, allowing the surface reconstructions of PtLuSb (001) films to be evaluated as a function of the termination procedure. $Al_{0.1}In_{0.9}Sb$ buffer layers were chosen due to the similarities between the zinc blende and half-Heusler ($C1_b$) crystal structures as well as the ability to lattice match the PtLuSb films[16]. Reflection high-energy electron diffraction (RHEED) was used to monitor film growth and paired with *in-situ* low-energy electron diffraction (LEED) to study the surface reconstructions of grown films. *Ex-situ* characterization included x-ray diffraction (XRD), to examine the bulk crystalline structure, and synchrotron x-ray photoemission spectroscopy (XPS) to probe the bulk and surface bonding of the films.

GaAs (001) substrates were used due to the availability of semi-insulating and highly doped wafers. Semi-insulating wafers were used for lateral transport measurements to limit parallel conduction paths while p+GaAs (001) wafers were used for photoemission measurements. Growth conditions were identical for each sample regardless of substrate choice. For growth, the GaAs wafers were loaded into a VG V80H III-V MBE system where the native oxide was desorbed under $As_4$ overpressure. A 500 nm buffer layer of GaAs was grown at a substrate temperature of 600 °C to form a smooth surface and to trap any remaining surface impurities after desorption of the native oxide. After the buffer layer growth, samples were annealed under an $As_4$ overpressure at 600 °C to form a well ordered (2x4)-$\beta$2 arsenic-terminated surface. Following the anneal, the substrate was cooled to 520 °C, at which point arsenic was shuttered. After a 10 minute anneal while residual $As_4$ was pumped out of the chamber, the substrate was further cooled to 380 °C. This cooling procedure retained the arsenic-stabilized (2x4)-$\beta$2 reconstruction at 380 °C. To nucleate $Al_{0.1}In_{0.9}Sb$, the antimony shutter was opened for a 10 second presoak of the GaAs surface followed by codeposition of $Al_{0.1}In_{0.9}Sb$. Typical group-V to group-III beam flux ratios were between 1.2 and 1.3 using a valved-cracker antimony source to supply an $Sb_2$ molecular beam. Due to the extremely high mismatch of 14.1 % between $Al_{0.1}In_{0.9}Sb$ and GaAs, the film immediately relaxes to form a three-dimensional structure, as observed by RHEED. Following nucleation, the substrate temperature was raised to 420 °C for continued growth. The higher growth temperature limited the sticking of excess antimony and improved adatom diffusivity. The RHEED pattern became streaky after approximately 15 nm of growth, indicating the coalescence and formation of a continuous, two-dimensional $Al_{0.1}In_{0.9}Sb$ surface. The $Al_{0.1}In_{0.9}Sb$ buffer was grown to either a thickness of 19.4 nm for lateral transport measurements or 200 nm for XRD and photoemission studies[17]. After growth, the $Al_{0.1}In_{0.9}Sb$ buffers were cooled under an $Sb_2$ overpressure and terminated with a mixed (1x3)/c(4x4) reconstruction before being transferred *in-situ* to a metals MBE chamber for PtLuSb growth.

For PtLuSb growth, elemental lutetium was evaporated from a high temperature effusion cell using a tantalum crucible, $Sb_4$ was evaporated from a conventional effusion cell using a pyrolytic-BN crucible, and platinum was evaporated from an e-beam evaporator. For lutetium and antimony, atomic fluxes were measured *in-situ* using a beam flux gauge with the collector connected to an electrometer for high precision current measurements. For platinum, atomic fluxes were monitored *in-situ* by a quartz-crystal microbalance located next to the sample. All *in-situ* flux measurements were correlated to actual atomic fluxes using Rutherford backscattering spectrometry to measure the atomic density of elemental films grown on silicon substrates at room temperature for set periods of time. These correlations were then used prior to each growth in order to set cell temperatures and e-beam emission currents to obtain the desired atomic flux ratios.

To promote the formation of PtLuSb, a shuttered growth technique was employed following the $C1_b$ PtLuSb crystal structure: the z = 0 plane of PtLuSb consists of 2 lutetium and 2 antimony atoms per unit cell and the z = ¼ plane consists of 2 atoms of platinum. Sequential deposition of $6.74 \times 10^{14}$ atoms/$cm^2$ of lutetium, followed by $6.74 \times 10^{14}$ atoms/$cm^2$ of platinum, and finally $6.74 \times 10^{14}$ atoms/$cm^2$ of antimony (one formula unit of PtLuSb) was repeated four times to form an 8 monolayer template at ~225°C. This epitaxial template was then used to grow thicker films of PtLuSb by codeposition of platinum, lutetium, and antimony at 320 – 380 °C. Pt:Lu:Sb flux ratios were set to 1:1:1.3. No evidence of excess antimony incorporation in the films at these growth conditions was observed in RHEED or XRD. This allowed for an antimony overpressure to be maintained during growth, similar to the growth of III-Sb semiconductors[18]. Typical growth rates were around 5 Å/min.

Films were capped with a protective 2 nm amorphous silicon, 5 nm $SiO_x$, or 100 nm antimony layer for *ex-situ* XRD, magneto-transport, or synchrotron XPS measurements, respectively, to prevent PtLuSb oxidation while in atmosphere. Carrier concentration and mobility was studied using by Hall measurements in the Van der Pauw geometry using indium contacts. To further understand the surface chemistry of the different surface reconstructions, synchrotron XPS measurements were taken as a function of incident photon energy at the i4 beamline at MAX-lab in Lund, Sweden. Synchrotron XPS measurement samples also included an additional 5 monolayer thick epitaxial GdSb diffusion barrier[19] between the $Al_{0.1}In_{0.9}Sb$ buffer layer and the PtLuSb film to minimize interfacial reactions during the thermal desorption of the sacrificial antimony cap. Complete thermal desorption of t he antimony cap was verified by LEED and core-level XPS. A (1x3) surface reconstruction, identical to that seen in the growth reconstruction study, was observed following de-cap. XPS spectra were fitted using a convolution of asymmetric Gaussian and Lorentzian line shapes. The full-width at half maximum for each component was held constant at 0.57 eV for 120 eV photon energies and 0.61 eV for 90 eV photon energies. Energy splitting between antimony $4d^{5/2}$ and antimony $4d^{3/2}$ was constrained to 1.25 eV for all fits.

## 3. Results and Discussion

### 3.1. Structural analysis of PtLuSb (001) thin films

Beginning with XRD measurements, Figure 1 shows the θ-2θ diffraction pattern for PtLuSb grown on a 200 nm $Al_{0.1}In_{0.9}Sb$ buffer on GaAs (001). The epitaxial relationship for the grown structures was PtLuSb (001) // $Al_{0.1}In_{0.9}Sb$ (001) // GaAs (001): PtLuSb [110] // $Al_{0.1}In_{0.9}Sb$ [110] // GaAs [110]. The absence of non-(00l) reflections confirms the growth of an epitaxial film with a (001) orientation. In addition, the absence of (00l) reflections, where l is odd, is consistent with the half-Heusler crystal structure. (002) and (004) reflections are also observed for the GaAs substrate and the $Al_{0.1}In_{0.9}Sb$ buffer layer. The out-of-plane lattice parameter of the PtLuSb film is measured to be 6.45 Å, aligned with the $Al_{0.1}In_{0.9}Sb$, and in good agreement with measurements of bulk crystals[10]. Further, we note the present of finite thickness fringes around both the (002) and the (004) peaks, suggesting the growth of abrupt, high quality interfaces. The film thickness calculated from the fringe spacing is 7.0 ± 0.1 nm, consistent with the expected growth rate as determined by beam flux calibrations.

### 3.2. Surface reconstructions of PtLuSb (001) thin films

Surfaces were characterized by RHEED and LEED as a function of substrate temperature and antimony overpressure. After post-growth annealing at various antimony overpressures and substrate temperatures, three unique surface phases were observed for PtLuSb (001): (1x3), c(2x2), and antimony capping. RHEED and LEED images of both the (1x3) and c(2x2) reconstructed surfaces show streaky patterns consistent with smooth high quality growth (Figure 2A). By varying both antimony overpressure and substrate temperature, a reconstruction phase diagram was obtained, shown in Figure 2B. (1x3) reconstructions are observed at low antimony overpressures or high substrate temperatures while c(2x2) reconstructions are observed for higher antimony overpressures or lower substrate temperatures. The stability of the c(2x2) reconstruction at lower substrate temperatures and higher antimony overpressures than the (1x3) reconstruction indicates that the c(2x2) surface reconstruction contains more antimony as compared to the (1x3) reconstruction. To retain (1x3) reconstructions at room temperature, the substrate temperature was held at approximately 360 °C while the antimony cell was shuttered, and then the sample was cooled to room temperature with no impinging antimony flux. c(2x2) reconstructions were retained at room temperature by cooling to 330 °C under an antimony overpressure of 8.75 x $10^{14}$ atoms/$cm^2$min until a c(2x2) reconstruction was obtained, and then shuttering antimony and cooling the substrate to room temperature. Further decreasing substrate temperature at constant antimony overpressure eventually results in the formation of elemental antimony on the surface.

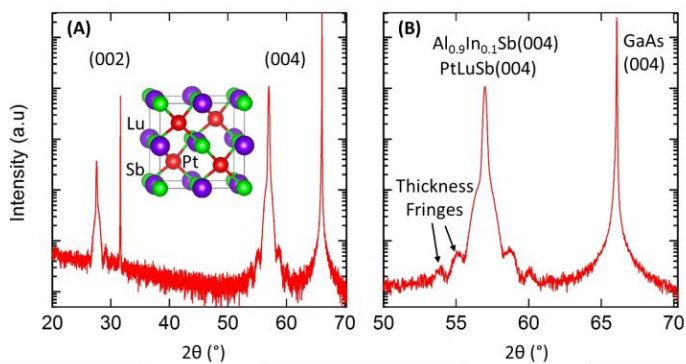

Figure 1. θ-2θ XRD scan showing the (002) and (004) reflections of the GaAs substrate, $Al_{0.1}In_{0.9}Sb$ buffer, and PtLuSb film. (A) Survey scan along (00l) direction. (B) Zoom in of the (004) reflection. The pattern confirms the growth of an epitaxial, (001)-oriented film. Finite thickness fringes correspond to a PtLuSb film thickness of 7.0 ± 0.1 nm.

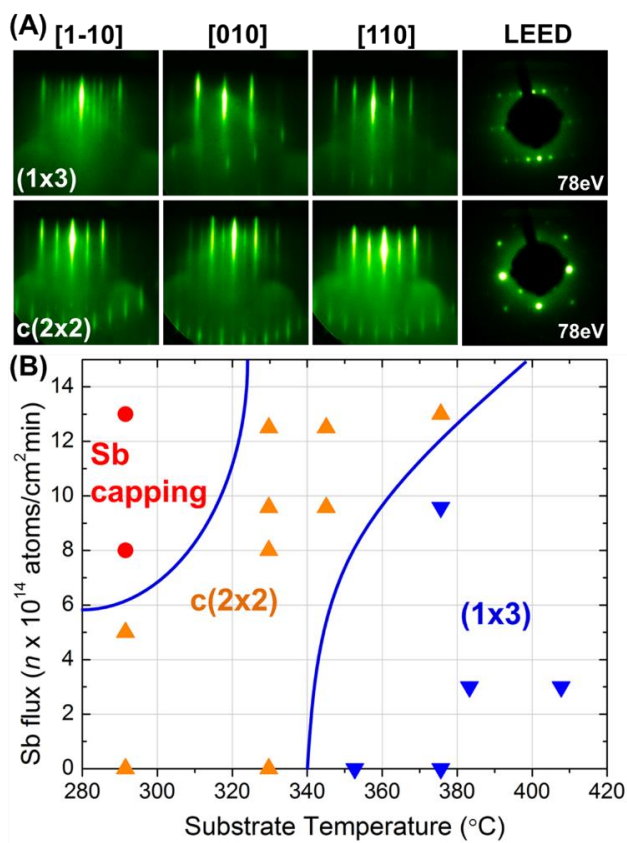

Figure 2. (A) RHEED and LEED images of (1x3) and c(2x2) reconstructed surfaces. (B) Reconstruction phase diagram for PtLuSb (001) as a function of antimony overpressure and substrate temperature. The presence of c(2x2) reconstructions at lower temperatures and higher antimony overpressures suggests that it is more antimony-rich than the (1x3) surface reconstruction.

### 3.3. X-ray photoemission spectroscopy of PtLuSb (001) thin films

The observation of surface reconstructions that change as a function of antimony overpressure or substrate temperature suggests a change in surface antimony composition for each reconstruction, similar to reconstruction changes observed in III-V semiconductors, like GaAs[20]. Figure 3 shows angle-integrated XPS spectra of the antimony 4d core level at incident photon energies of 120 eV (less surface sensitive) and 90 eV (more surface sensitive) for (1x3) reconstructed PtLuSb (001). Two components, S1 and S2, are observed at lower binding energy ($\Delta BE_{S1}$= -0.46 ± 0.03 eV, $\Delta BE_{S2}$= -0.92 ± 0.03 eV) than the bulk component (bulk antimony $4d^{5/2}$ = 32.21 ± 0.02 eV). The peak ratio of the components S1 and S2 ($A^{S1}/A^{S2}$) remains nearly constant at 1.1 for both photon energies. However, the total component to bulk ratio ($A^{S1+S2}/A^{bulk}$) at $hv$ = 90 eV is 2.1 as compared to 1.4 at $hv$ = 120 eV. Since the hv = 90 eV measurement is more surface sensitive, the larger total component to bulk ratio confirms that S1 and S2 arise from bonding at the surface. The S1 component, with a -0.46 eV shift from the bulk component, is analogous to surface components observed in III-V semiconductors like GaAs (001) and GaSb (001) where a similarly shifted components have been attributed to group-V dimerization at the surface[21,22]. The S2 component, which is shifted -0.92eV from the bulk component, represents a more negative bonding environment for antimony, and could be attributed to broken antimony dimers leading to unpaired electrons at step edges or other surface defects. Notably, both antimony core-level shifts towards lower binding energies provide compelling evidence to indicate that antimony-antimony dimers are present in the (1x3) surface reconstruction. Antimony de-capping limitations prevented similar analysis with the c(2x2) surface reconstruction, although the reconstruction phase diagram

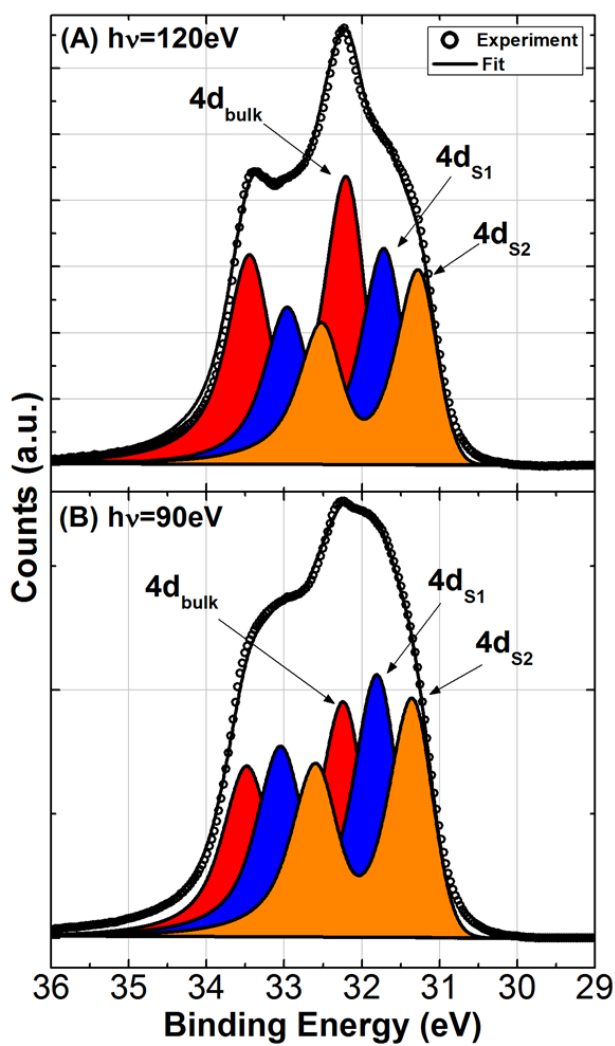

*Figure 3. XPS measurements of the antimony 4d core level as a function of incident photon energy. At (A) 120 eV photon energy, three components are observed: bulk, S1, and S2. The more surface sensitive (B) 90 eV photon energy scan shows an increase in peak area of components S1 and S2 as compared to the bulk component, indicating the surface nature of these components.*

suggests that dimers or an even further antimony saturated surface should be present.

### 3.4. Electrical transport of PtLuSb (001) thin films

The reconstruction phase diagram presented in Figure 2B highlights a growth window of substrate temperatures between 320 – 410 °C where an antimony flux of 8.75 x $10^{14}$ atoms/$cm^2$min results in a stable antimony-terminated surface without the formation of an elemental antimony capping layer or the appearance of additional features in the RHEED/XRD patterns. This verifies the existence of a growth window where antimony overpressures can be used while still retaining approximately 1:1:1 Pt:Lu:Sb stoichiometry in the film. Within this growth window, substrate temperature was found to play a large role in electronic film quality, as measured by room temperature carrier concentration and mobility (Figure 4). A clear trend showing decreasing carrier concentration as well as increasing mobility is observed as growth temperature is increased. Both behaviors can be attributed to the reduction of point defects, which lead to the introduction of carriers and as well as charged impurities that may serve as scattering centers, further reducing mobility. The observed trends in carrier concentration and mobility point towards growth at a higher substrate temperature, which would increase adatom mobility as well as reduce the incorporation of non-stoichiometric antimony. Unfortunately, above growth temperatures of 380 °C, the PtLuSb half-Heusler phase is not stable on $Al_xIn_{1-x}Sb$, and significant film-substrate reactions begin to occur. While these reactions provide an upper limit for growth temperature in the PtLuSb/$Al_xIn_{1-x}Sb$ system, it is possible that another substrate could be used that is less reactive with PtLuSb, enabling growth at higher substrates temperatures, potentially resulting in higher quality films. Other studies have used MgO (001) substrates with metallic buffers, such a Ta (001), for higher temperature growths[23], but these layers are conductive, making it difficult to study the electronic structure of the film by lateral transport measurements and limiting access to the helical surface states. Furthermore, the higher symmetry of these metal layers allows for the formation of additional rotational and antiphase defects within the Heusler compound which is undesirable.

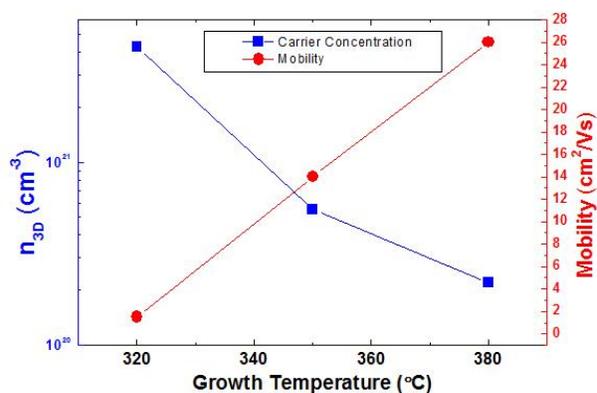

*Figure 4. Room temperature Hall measurements in the Van der Pauw geometry showing trends of decreasing carrier concentration and increasing mobility with increasing growth temperature. This behavior is attributed the decrease in point defect density and/or an increase in crystal quality at higher growth temperatures.*

## 4. Conclusions

This work demonstrates the optimization of the growth of epitaxial PtLuSb (001) thin films on $Al_{0.1}In_{0.9}Sb$ (001)/GaAs (001) heterostructures. The observation of both (1x3) and c(2x2) surface reconstructions under an antimony overpressure indicated that these surface reconstructions results from differences in the surface bonding of antimony. XPS analysis suggests some degree of dimerization exists between the surface antimony atoms as evidenced by the observation of energy shifts towards lower binding energy in antimony 4d surface components. The stability of the surface reconstructions as a function of substrate temperature and antimony overpressure led to the identification of a growth window were an antimony overpressure could be utilized. For Pt:Lu:Sb atomic flux ratios of 1:1:1.3, a substrate temperature of 320 – 380 °C lead to stoichiometric films with the best electrical properties obtained near 380 °C. Further increased growth temperatures lead to film-substrate reactions between the PtLuSb and the $Al_xIn_{1-x}Sb$ buffers used in this study, but may result in further improvements in film quality if a more stable epitaxial template can be identified. The use of surface reconstructions to determine a growth window is applicable to the epitaxial growth of other half-Heusler compounds comprised of group-V elements and should enhance their ability to be integrated with various III-V heterostructures.

**Acknowledgements**

The authors would like to acknowledge the support of the beamline staff and scientists at the i4 beamline in MAX-LAB. Beamline measurements were partially supported by the IMI Program of the National Science Foundation under Award No. DMR 08-43934. X-ray diffraction measurements were taken at a shared MRSEC funded facility under award number NSF-DMR-1121053. Funding for this work was provided by the Office of Naval Research under award number N00014-11-1-0728 as well as the Army Research Office under award number W911NF-12-1-0459.